\documentclass[usenatbib]{mn2e}
\usepackage{graphicx}
\usepackage{pstricks}
\usepackage{textcomp}
\usepackage{pst-plot}

\date{Accepted 2010 January 18}

\newlength{\figurewidth}
\title[Two step ejection of massive stars]{Two step ejection of massive stars and the issue of their formation in isolation}
\author[J.~Pflamm-Altenburg]
{Jan~Pflamm-Altenburg\thanks{email: jpflamm@astro.uni-bonn.de,
    pavel@astro.uni-bonn.de}
  and Pavel~Kroupa\footnotemark[1]\\
  Argelander-Institut f\"ur Astronomie (AIfA), University of Bonn, 
  Auf dem H\"ugel 71, D-53121 Bonn, Germany \\
}
\begin{document}
\maketitle
\begin{abstract}
  In this paper we investigate the combined effect of massive binary ejection
  from star clusters and a second acceleration of a massive star 
  during a subsequent supernova
  explosion. 
  We call this the \emph{two-step-ejection} scenario.
  The main results are: i) Massive field stars produced via the 
  two-step-ejection process can not in the vast majority of cases
  be traced back to their parent star 
  clusters. These stars can be mistakenly considered as massive stars formed 
  in isolation. ii) The expected O star fraction produced via 
  the two-step-ejection process is of the order of 1--4~per~cent, in 
  quantitative agreement with the observed fraction of candidates for 
  isolated-O-star formation. 
  iii) Stars ejected 
  via the two-step-ejection process can
  get a higher final velocity (up to 1.5--2 times higher) 
  than the pre-supernova  velocity of the massive-star binary.
\end{abstract}
\section{Introduction}
Considering pure number counts massive stars are by far only a tiny minority
in the stellar population of galaxies. But they mainly drive 
galactic evolution due to their dominating chemical and energetic feedback.
Although  the importance of massive stars
for galactic astrophysics has been accepted, the 
physical circumstances of their formation are still not resolved, i.e. where,
why, and how they form.

It is currently strongly debated whether the formation of massive stars is
entirely restricted to the interior of massive star clusters or if they can form
in isolation  in the galactic field. 

Indeed, on the basis  of a statistical analysis of a sample of 
galactic O stars \citet{dewit2005a} conclude
that 4$\pm$2 per~cent of all O-type stars can be considered 
as formed outside a cluster environment.
They further show that this fraction of isolated O stars is expected
if the slope of the cluster mass function (CMF) is $\beta=1.7$.
This assumed CMF slope in the low mass star cluster regime is in 
disagreement with the slope of $\beta=2$ observed
in  the solar neighborhood \citep{lada2003a}.

The definition of an isolated O star in the IMF-Monte-Carlo simulations
of \citet{dewit2005a} is restricted to stellar ensembles which contain 
only one O star.
\citet{parker2007a} strengthened the definition of an isolated O star
being an O star without B-star companions  but allowed the O star to be 
surrounded by a  cluster with a mass of $<$100~M$_\odot$. 
Using this definition \citet{parker2007a}
conclude that an observed CMF slope of $\beta=2$ can quantitatively 
explain the statistical analysis by \citet{dewit2005a} 

The analysis by \citet{dewit2004a,dewit2005a}
identifies a small number of O-stars which are deemed
to be truly isolated in the sense of not being traceable to an origin
in a cluster or OB association. 
However, \citet{gvaramadze2008a} reported
the existence of a bow shock
associated with the O-star HD165319, which is marked in \citet{dewit2005a}
as a very likely candidate for an O-star formed in isolation. 

In general, two main processes exist for the 
production of high-velocity O- and B-stars: i)
close encounters between binaries and single stars or binaries and binaries
can result in the ejection of massive stars 
\citep[e.g.][]{poveda1967a,hoffer1983a,mikkola1983a,mikkola1984a,leonard1991a}.
The ejection velocity is of the order of the orbital velocity of the binary
\citep{heggie1980a}. Thus tighter binaries can produce larger ejection 
velocities. ii) The  supernova explosion of one component of the binary
leads to a recoil of the other component
\citep[e.g.][]{zwicky1957a,blaauw1961a,iben1997a,tauris1998a,portegies_zwart2000a,hoogerwerf2001a}. 
The supernova ejection scenario only requires the existence of massive binaries,
whereas ejection rates in the dynamical ejection scenario depend on the
close encounter frequency. This frequency will be increased if massive 
stars form in 
compact few-body groups \citep{clarke1992a,pflamm-altenburg2006a}.

Various studies on the individual ejection processes exist.
To our knowledge only one investigation exists combining both 
processes \citep*{gvaramadze2008b}, where the hypothesis is explored
that  a hyperfast pulsar  can be the remnant of a symmetric supernova explosion 
of  a massive O star dynamically ejected from  a young massive star cluster.
In this contribution
we investigate for the first time the combination of
the dynamical and supernova ejection process, in which a massive binary is 
dynamically ejected from a star cluster with subsequent supernova explosion 
of one binary component with recoil of  the other binary component.
We refer to
this composite ejection scenario
as the \emph{two-step-ejection process} of massive stars.

We start our investigation in Section~\ref{sec_comp_speed}
with the calculation of the velocity spectrum 
of stars which are released with the same velocity during a supernova
from binaries with identical ejection velocities
We then derive the probability that stars released by
a supernova from ejected binaries
can be traced back to their parent star cluster 
(Section~\ref{sec_back_trace_prob}), and discuss
the maximum possible velocity which stars 
can get in the two-step-ejection process
 in Section~\ref{sec_max_speed}.

\section{Compound velocity spectrum}
\label{sec_comp_speed}
Due to dynamical interactions during close encounters
of stars, binaries can be ejected from star clusters with the ejection
velocity, $v_\mathrm{e}$. 
If one component of the binary explodes in a supernova the gravitational force acting on the other component decreases rapidly if the expanding
supernova shell has passed the orbit of the stellar companion releasing it with
nearly its orbital velocity into the galactic field. Depending on the configuration of the pre-supernova binary and the supernova details (e.g. eccentricity, 
supernova mass loss, asymmetry of the supernova) the supernova remnant 
may still be bound to its companion. \citet{portegies_zwart2000a} calculated
that between 20 and 40~per~cent of the supernova runaways have neutron star 
companions, but less than 1~per~cent are detectable as pulsars.
Independent if the supernova remnant (neutron star or black hole)
still remains bound to the other massive star or not we speak of 
binary disintegration throughout this paper. The post-supernova 
massive-star runaway is released with a release velocity, $v_\mathrm{r}$, with respect to the centre-of-mass of the pre-supernova binary.

The vectorial
sum of both velocities, the ejection and release velocity, is the new
velocity of the released star. We call this velocity the compound
velocity, $v_\mathrm{c}$. For fixed ejection velocity, $v_\mathrm{e}$, and
release velocity, $v_\mathrm{r}$, the compound velocity, $v_\mathrm{c}$,
is distributed, because the direction of the release of the star from 
the disintegrating binary during the supernova is
randomly distributed in space. The corresponding spectrum of the 
compound velocity, $f_\mathrm{c}(v_\mathrm{c})$, defines the
number of stars, $dN(v_\mathrm{c})$, which have a compound velocity, 
$v_\mathrm{c}$, after they have been released from ejected binaries.

\subsection{Calculating $f_\mathrm{c}(v_\mathrm{c})$}
We start the calculation of the resulting compound velocity spectrum
with the definition of the compound velocity through vectorial addition,
\begin{equation}
\bmath{v}_\mathrm{c}=\bmath{v}_\mathrm{e}+\bmath{v}_\mathrm{r}\;.
\end{equation}
\begin{figure}
  \begin{center}
    \includegraphics[width=\figurewidth]{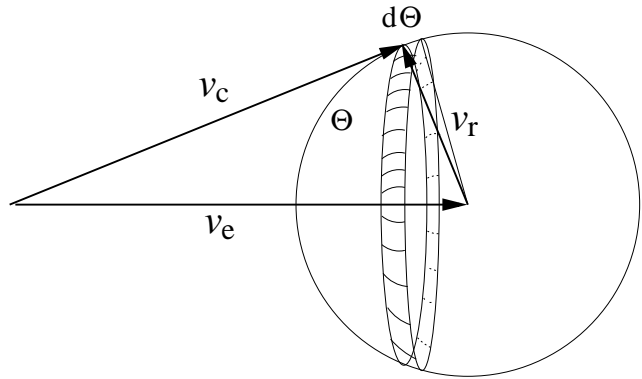}
  \end{center}
\caption{Illustration of the compound velocity. A binary is ejected
with the velocity $v_\mathrm{e}$. One component is then released with
the velocity $v_\mathrm{r}$ resulting in the compound velocity,
$v_\mathrm{c}$, given by the vectorial sum.}
\label{fig_sketch}
\end{figure}   
The relation between the absolute values of the 
velocities follows from the cosine theorem (Fig. \ref{fig_sketch}),
\begin{equation}
v_\mathrm{c}^2 = v_\mathrm{e}^2 + v_\mathrm{r}^2 
- 2v_\mathrm{e}v_\mathrm{r}\cos{\theta}\;.
\end{equation}
Differentiation leads to the relation 
\begin{equation}
\frac{d v_\mathrm{c}}{d\theta} = 
\frac{v_\mathrm{e}v_\mathrm{r}}{v_\mathrm{c}}\sin{\theta}
\end{equation}
between the compound velocity and the release angle $\theta$.

The orientations of the binaries are randomly distributed in space. Thus,
the released stars are 4$\pi$ distributed with respect the centre-of-mass
system of the binary. If a set of binaries release $N$ stars isotropically
the number of stars $d N(\theta)$ released in a small angle $d\theta$
is given by the ratio of the area 
of the small circular stripe, $d A(\theta)$,  with  the angle $\theta$
and the unit sphere, $4\pi$ (Fig.~\ref{fig_sketch}) , 
\begin{equation}
  \frac{d N(\theta)}{N} = \frac{d A(\theta)}{4\pi}\;,
\end{equation}
where the area of the small circular stripe is 
\begin{equation}
d A(\theta) = 2\pi \sin(\theta) d \theta.
\end{equation}
By combining these equations the number fraction of stars having
the compound velocity $v_\mathrm{c}$ is
\begin{equation}
\frac{d N(v_\mathrm{c})}{d v_\mathrm{c}} = 
\frac{N}{2 v_\mathrm{e} v_\mathrm{r}}v_\mathrm{c}\;.
\end{equation}
As the angle $\theta$ varies from 0 to $2 \pi$, the allowed range of the
compound velocity can be obtained from the cosine theorem above and
is
\begin{equation}
|v_\mathrm{e}-v_\mathrm{r}| \le v_\mathrm{c}\le v_\mathrm{e}+v_\mathrm{r}\;.
\end{equation}

The distribution function, $f_\mathrm{c}(v_\mathrm{c})$,  
of the compound velocity is normalised by
\begin{equation}
  \label{eq_f_norm}
  1 = \int f_\mathrm{c}(v_\mathrm{c})\;dv_\mathrm{c}\;,
\end{equation}
and is for a constant ejection velocity, $v_\mathrm{e}$,
and constant release velocity, $v_\mathrm{r}$, 
\begin{equation}\label{eq_f_c}
f_\mathrm{c}(v_\mathrm{c}) = \frac{1}{2 v_\mathrm{e} v_\mathrm{r}}v_\mathrm{c}
\Theta(v_\mathrm{c} - |v_\mathrm{e}-v_\mathrm{r}|) 
\Theta(v_\mathrm{e}+v_\mathrm{r} - v_\mathrm{c})\;,
\end{equation}
where the $\Theta$-mapping is defined by
\begin{equation}
\Theta(x) = \left\{
\begin{array}{l@{\;\;\;;\;}c}
1 & x\ge 0\\
0 & x<   0
\end{array}
\right.\;.
\end{equation}
\begin{figure}
  \begin{center}
    \includegraphics[width=\figurewidth]{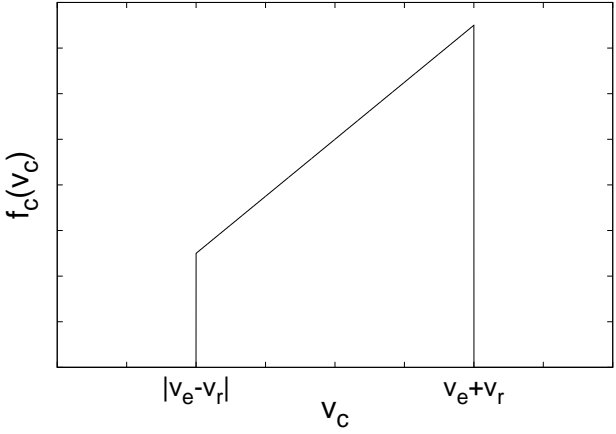}
    \end{center}
\caption{The form of the compound velocity spectrum $f_\mathrm{c}$ 
  (eq.~\ref{eq_f_c}).}
\label{compund_velocity_spectrum}
\end{figure}
The form of the compound spectrum can be seen in 
Fig. \ref{compund_velocity_spectrum}. Compound velocities
at the high speed end are  preferred. But the distribution 
function, $f_\mathrm{c}(v_\mathrm{c})$, 
flattens with increasing ejection or release velocity.
The distribution is symmetric in $v_\mathrm{r}$ and $v_\mathrm{e}$.
Note that in eq.~\ref{eq_f_norm} an additional factor $4\pi\;v_\mathrm{c}^2$, 
as for example 
in the Maxwellian distribution calculated a from three-dimensional Gaussian distribution function, is not required, because 
$f_\mathrm{c}(v_\mathrm{c})$
refers already to an absolute velocity value and is not calculated from 
spatial integration over a three-dimensional distribution function. 

If the velocities of the ejected binaries and of the released stars 
are distributed according to uncorrelated 
distribution functions, $f_\mathrm{e}(v_\mathrm{e})$
and $f_\mathrm{r}(v_\mathrm{r})$, then the resulting distribution 
of compound velocities is calculated by integration over both 
distributions,
\begin{eqnarray}
  f_\mathrm{c}(v_\mathrm{c})&=&
\frac{v_\mathrm{c}}{2}\int\int
\frac{f_\mathrm{e}(v_\mathrm{e})f_\mathrm{r}(v_\mathrm{r})}{v_\mathrm{e}v_\mathrm{r}}
\nonumber\\
&&\Theta(v_\mathrm{c} - |v_\mathrm{e}-v_\mathrm{r}|) 
\;\Theta(v_\mathrm{e}+v_\mathrm{r} - v_\mathrm{c})\;
dv_\mathrm{e}\;dv_\mathrm{r}
\;.
\end{eqnarray} 

\subsection{Properties of $f_\mathrm{c}(v_\mathrm{c})$}
\begin{figure}
  \begin{center}
    \includegraphics[width=\figurewidth]{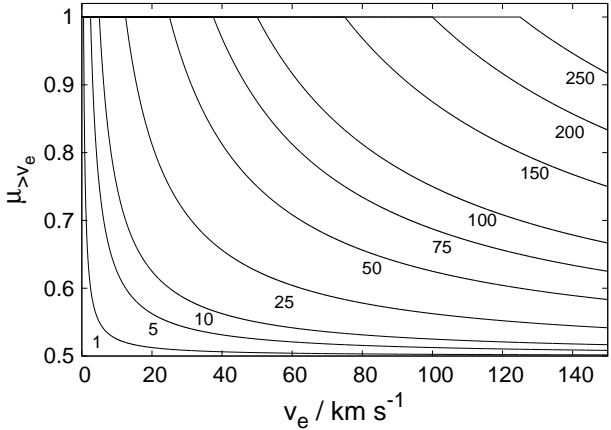}
  \end{center}
  \caption{The fraction of accelerated stars in dependence of the 
    ejection and release velocity for different release velocities
    in km~s$^{-1}$. }
  \label{acceleration_fraction}
\end{figure}

In the following we derive some properties of the compound
velocity spectrum.

i) The simplest case we can consider is that if one of the velocities, 
$v_\mathrm{e}$ or $v_\mathrm{r}$, is zero.
If for example $v_\mathrm{r}$ converges against zero then the velocity 
spectrum converges against the delta-distribution 
\begin{equation}
f_\mathrm{c}(v_\mathrm{c}) = 
\lim_{v_\mathrm{r} \to 0}f_{v_\mathrm{r}}(v_\mathrm{c}) = 
\delta(v_\mathrm{c}-v_\mathrm{e})\;.
\end{equation}
Because $f_\mathrm{c}(v_\mathrm{c})$ 
is symmetric in $v_\mathrm{r}$ and $v_\mathrm{e}$,
the same result follows for $v_\mathrm{e}\to 0$.
The compound velocity is identical to the ejection or release velocity.

ii) The released stars are not necessarily faster than the previous binaries.
They can also be decelerated. 
The fraction of stars ($\mu_{>v_\mathrm{e}}$) which are accelerated, i.e.
having a compound velocity greater than the ejection velocity
is given by the integral
\begin{equation}
\mu_{>v_\mathrm{e}} = \int_{v_\mathrm{e}}^{v_\mathrm{e}+v_\mathrm{r}}
f_\mathrm{c}(v_\mathrm{c})\;\mathrm{d}v_\mathrm{c}\;.
\end{equation}
The evaluation of the integral leads to
\begin{eqnarray}
\mu_{>v_\mathrm{e}} &=& \frac{1}{4v_\mathrm{e}v_\mathrm{r}}
\left(v_\mathrm{r}^2+2v_\mathrm{e}v_\mathrm{r}-\right.\nonumber\\
&&\left.\Theta(|v_\mathrm{e}-v_\mathrm{r}|-v_\mathrm{e})
\left(v_\mathrm{r}^2-2v_\mathrm{e}v_\mathrm{r}\right)
\right)
\;.
\end{eqnarray}
The fraction of accelerated stars in dependence of the ejection and
release velocity can be seen in Fig. \ref{acceleration_fraction}.
Two cases can be distinguished:
\begin{equation}
|v_\mathrm{e}-v_\mathrm{r}|\ge v_\mathrm{e}\;:\;
\mu_{>v_\mathrm{e}}=1\;,
\end{equation}
and
\begin{equation}
|v_\mathrm{e}-v_\mathrm{r}|< v_\mathrm{e}\;:\;
\mu_{>v_\mathrm{e}}=\frac{v_\mathrm{r}}{4\;v_\mathrm{e}}+\frac{1}{2}\;.
\end{equation}
The number fraction of accelerated stars is always larger than 50~per cent and
all stars are accelerated for $v_\mathrm{r}>2\;v_\mathrm{e}$.

For the case that $v_\mathrm{e}=v_\mathrm{r}$, which will be important for 
Section~\ref{sec_max_speed}, i.e. the ejection velocity is comparable to the
orbital velocity of the ejected binary,
 the fraction of accelerated stars is 75~per~cent.

iii) We  now calculate the resulting mean compound velocity,
$\bar{v}_\mathrm{c}$, by the integral
\begin{equation}
\bar{v}_\mathrm{c} =
\int_{|v_\mathrm{e}-v_\mathrm{r}|}^{v_\mathrm{e}+v_\mathrm{r}}
v_\mathrm{c}f(v_\mathrm{c}) \mathrm{d}v_\mathrm{c} = 
\frac{1}{6 v_\mathrm{e}v_\mathrm{r}} 
\left. v_\mathrm{c}^3\right|_{|v_\mathrm{e}-v_\mathrm{r}|}^{v_\mathrm{e}+v_\mathrm{r}}
\end{equation}
We define
\begin{equation}
v_\mathrm{min} = \min\{v_\mathrm{e},v_\mathrm{r}\}
\;\;\;,\;\;\;
v_\mathrm{max} = \max\{v_\mathrm{e},v_\mathrm{r}\}
\end{equation}
and write
\begin{equation}
v_\mathrm{e}+v_\mathrm{r} = v_\mathrm{min} + v_\mathrm{max}
\end{equation}
and
\begin{equation}
|v_\mathrm{e}-v_\mathrm{r}| = v_\mathrm{max} - v_\mathrm{min}\;.
\end{equation}
and finally the  mean compound velocity can be written as
\begin{equation}
\bar{v}_\mathrm{c} = \frac{1}{3}
\;\frac{v_\mathrm{min}^2+3 v_\mathrm{max}^2}{v_\mathrm{max}}\;.
\end{equation}
It follows that the mean compound velocity is always greater than
or equal to the ejection and release velocity:
\begin{equation}
  \label{eq_mean_v}
  \bar{v}_\mathrm{c} \ge \frac{1}{3}\;\frac{3\;v_\mathrm{max}^2}{v_\mathrm{max}}
  = v_\mathrm{max}\;.
\end{equation}

For the case that $v_\mathrm{e}=v_\mathrm{r}$, which will be important for 
section~\ref{sec_max_speed}, i.e. the ejection velocity is comparable to the
orbital velocity of the ejected binary,
the mean compound velocity is 
\begin{equation}
  \bar v_\mathrm{c} = \frac{4}{3}\;v_\mathrm{e}
\end{equation}

\section{Back-tracing probability}
\label{sec_back_trace_prob}
If single stars or binaries are ejected from star clusters
it is theoretically possible to calculate their orbits backward,
if the force field, through which the objects have moved in time, 
is given.
In such a  case the star cluster, where the stars have their origin, 
can be identified. 

\begin{figure}
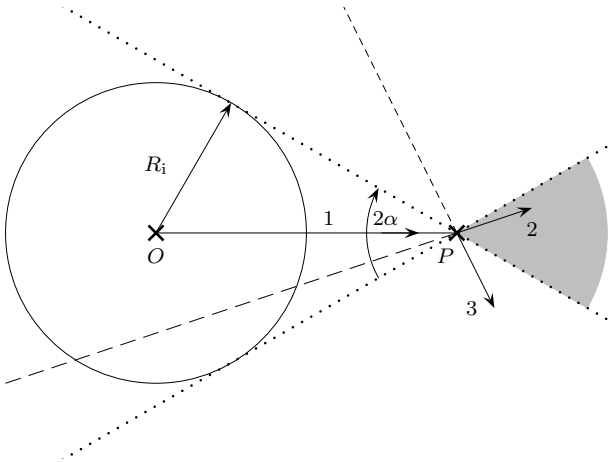

  \begin{center}
    \SpecialCoor
    \psset{linewidth=0.01pt}
    \psset{arrowsize=4pt 3}
    \pspicture*(-2,-3)(6.2,3)
    \pswedge[linecolor=lightgray,fillstyle=solid,fillcolor=lightgray](4,0){2.0}{-29.92}{29.92}
    \pscircle(0,0){2}
    \psline{->}(0,0)(2;60)
    \rput(0,0.9){$R_\mathrm{i}$}
    \psline[linewidth=1pt](-0.1,0.1)(0.1,-0.1)
    \psline[linewidth=1pt](0.1,0.1)(-0.1,-0.1)
    \rput(0,-0.3){$O$}
    \psline[linewidth=1pt](3.9,0.1)(4.1,-0.1)
    \psline[linewidth=1pt](4.1,0.1)(3.9,-0.1)
    \rput(3.85,-0.3){$P$}
    \psline(0,0)(4,0)
    \psline{->}(3,0)(3.5,0)
    \psline{->}(4,0)(5,0.333333333333)
    \rput(2.3,+0.2){1}
    \psline[linestyle=dashed,dash=6pt 3pt](4,0)(-2.0,-2.0)
    \rput(5.0,0.05){2}
    \psline{->}(4,0)(4.5,-1.0)
    \psline[linestyle=dashed,dash=3pt 2pt](4,0)(2,4)
    \rput(3.05,0.2){2$\alpha$}
    \psarc{<-}(4,0){1.2}{150}{210}
    \rput(4.2,-1.0){3}
    \psline[linestyle=dotted,linewidth=1pt](6.0,-1.151079137)(-2.95,4)
    \psline[linestyle=dotted,linewidth=1pt](6.0,1.151079137)(-2.95,-4)
    
    \endpspicture
    \end{center}
    \caption{Illustration of the back-tracing-probability. See text for details.}
    \label{scetch_back_tracing}
\end{figure}

If the binary disintegrates due to a supernova, 
the released component suffers a strong deflection
of its previous binary orbit, and
the cluster, from where it is expelled, can only be identified if
the angle between the deflected and previous orbit is not too large
(Fig. \ref{scetch_back_tracing}).

A binary is ejected from a star cluster and moves along path 1.
After the binary has moved the distance $\xi$ from the centre of the 
cluster, $O$, it disintegrates at the position $P$ 
and one component of the binary is released.
The star cluster, from which the star has been ejected, is identified, 
if the extrapolated path of the released star intersects a sphere
with the identification radius, $R_\mathrm{i}$, round the star cluster.
If the released star moves along path 2, the star cluster is
identified (extrapolated long dashed line). 
If the star moves along path 3, the parent star cluster
can not been identified (extrapolated short dashed line).
Only stars which have new orbits within the gray shaded region between 
the dotted lines can be traced back
to their parent star cluster.
  
At the location of binary disintegration (point $P$)
the identification sphere appears under the angle $2 \alpha$, i.e.
a solid angle of $2 \pi (1-\cos\alpha)$. 
Because the directions of the released stars are randomly distributed over the 
full solid angle of $4 \pi$, the probability that the star cluster can be
identified is given by the ratio of the solid angle of the identification
sphere and the full solid angle,
\begin{equation}
P = \frac{1-\cos\alpha}{2}\;\;\;.
\end{equation}
The cosine  can be expressed by the identification radius, $R_\mathrm{i}$,
and the disintegration distance, $\xi$,
\begin{equation}
\cos\alpha = \sqrt{1-\sin^2\alpha} = \sqrt{1-\frac{R_\mathrm{i}^2}{\xi^2}}\;\;\;.
\end{equation}      
Then the identification probability is
\begin{equation}
P = \frac{1}{2}\left(1-\sqrt{1-R_\mathrm{i}^2/\xi^2}
\right)\;\;\;.
\end{equation}
The resulting probabilities for different ratios of the identification
radius and disintegration distance are listed in Table~\ref{tab_probability}.

\begin{table}
\caption{Back-trace probability}
\begin{tabular}{ccccccc}
\hline
$R_\mathrm{i}$/$\xi$& 1  & 1/2 &1/5  &1/10  &  1/15 &1/50   \\
$P$ [\textperthousand]           & 500 & 67 &10   & 2.5 & 1 & 0.1\\ 
\hline
\end{tabular}

\medskip
$R_\mathrm{i}$ is the radius of the identification sphere
around the star cluster from which the binary has been
ejected. $\xi$ is the distance between the location of the
disintegration of the binary and the centre of the
identification sphere which coincides with the centre of
the star cluster. $P$ is the back-tracing probability in
\textperthousand. 
\label{tab_probability}
\end{table}

\section{Observed statistics of runaways and apparently isolated O stars}
Massive stars and massive binaries can only be ejected dynamically
from star clusters during the early stage of their life. If they form
within compact few-body configurations or trapezia systems \citep{clarke1992a}, 
then the decay time scale of these few-body systems ($<$1~Myr, 
\citealp{pflamm-altenburg2006a}) implies only
early ejections of massive stars. If massive stars and binaries are 
formed distributed over the  star cluster, then ejections can only
occur as long as the stellar densities are high. Due to gas expulsion 
young embedded star clusters become super-virial and start to expand and
the stellar density decreases rapidly \citep*{kroupa2001b}. The time scale
of decrease of the stellar density is comparable to the gas expulsion time
scale, of the order of $\approx$1~Myr.
Thus the time of flight of ejected binaries
is comparable to their maximum-lifetime.

Taking a lower velocity cut-off for O-star runaways of 30~km~s$^{-1}$ as
considered in \citet{gies1986a} and a mean life-time of 5~Myr of O-stars
implies a disintegration distance $\xi=$150~pc. For a large identification 
radius of $R_\mathrm{i}=$10~pc of the star cluster, the back-tracing probability
is  1~\textperthousand (Table~\ref{tab_probability}). Lowering the cut-off
velocity of the runaway-star definition to 10~km~s$^{-1}$ results in a 
back-tracing probability of 1~per~cent. Thus, ejected massive binaries which
are listed in runaway O-star surveys will produce O stars which can not be 
traced back to their parent star cluster. But one might expect that in this
case the massive star can be traced back to a supernova shell. 
However, as single supernova shells disappear on a time scale of 0.5--1~Myr 
\citep{chevalier1974a} it might be possible to identify the parent supernova
of the released star only in very rare cases.

The observationally derived  
runaway fraction among O stars varies widely in the literature,
(see for example Table~13 in \citealp{gies1986a} or Table~5 
in \citealp{stone1991a}). \citet{gies1986a} identified 15 stars out of a 
sample of 36 runaway candidates with confirmed peculiar radial velocities 
$\ge$30~km~s$^{-1}$. Comparing with their stated total number of about 
90~O stars within
their sample space, a runaway fraction of 15/90= 16~per~cent results.
They further conclude that the true runaway fraction of O stars depends
on the adopted velocity cut-off and may lie in the range of 10--25~per~cent.
\citet{gies1986a} also found the binary fraction among runaway O stars in their 
sample to be about 10~per~cent. As explained above, O stars released in a
supernova in an ejected massive binary result in field O stars which can not 
be traced back to their parent star clusters. Thus, 
on the basis of the O-star runaway
fraction and binary fraction of O-runaway data published by \citet{gies1986a}
1--2.5~per~cent of O stars can not be traced back
to the star cluster where they have form. These O stars will appear to have
formed in isolation, although they were born in an ordinary star cluster.

The different O-star runaway fractions in the literature have been unified by 
\citet{stone1991a} by considering true space frequencies:
The radial velocity spectrum of O stars is decomposed into two different
Gaussian velocity distributions, which correspond to two different
Maxwellian space velocity distributions. The high velocity component 
has a number fraction of $f_\mathrm{H}$=46~per~cent and a velocity distribution
of $\sigma_\mathrm{H}$=28.2~km~s$^{-1}$. 
By transforming individual O star 
runaway studies, which are based on individual 
runaway definitions, to true space frequencies
based on bimodality in the velocity distribution of O stars,
\citet{stone1991a} achieves good agreement between the different
individual studies. 

From the runaway fraction of 46~per~cent derived by \citet{stone1991a}
it follows that a fraction of 4.6~per~cent of O stars 
will apparently form in isolation,
if the runaway binary fraction of 10~per~cent by \citet{gies1986a} is used.

Thus, the two-step-ejection process predicts a fraction of O stars which
have formed apparently in isolation in the range of 1--4.6~per~cent.
\citet{dewit2004a,dewit2005a} conclude, based on the actually observed
positions and velocities of O stars, that 4$\pm$2~per~cent of O stars
can be considered as candidates of massive stars formed in isolation,
because such stars can not be traced back to a young star cluster.
Consequently, the process of two-step ejection can quantitatively account 
for the proposed fraction of massive candidates formed in isolation.

\section{Maximum possible velocity}
\label{sec_max_speed}
The maximum possible velocity in the two-step-ejection process is 
$v_\mathrm{max}=v_\mathrm{r}+v_\mathrm{e}$, i.e. if the star is released 
in the moving direction of the previous binary. Binaries can be ejected from 
star clusters due to dynamical interactions. Two common situations are the
scattering of two binaries (B+B) and the scattering of one binary and one single
star (B+S). B+B-events lead commonly to two single runaway stars and one
tight binary. But also high velocity binaries can be produced in roughly 
10~per~cent of all cases \citep{mikkola1983a}. 

In the B+S event
high velocity binaries must be produced due to local conservation of linear 
momentum. The ejection velocity of the single star is typically of the
order of the orbital velocity of the binary 
\citep{heggie1980a}, $v_\mathrm{o}$. In the case of
equal masses the ejection velocity of the binary, $v_\mathrm{e}$, 
is half of the ejection velocity of the single star, 
$v_\mathrm{e}=\frac{1}{2}\;v_\mathrm{o}$.  When the ejected binary 
disintegrates due to a supernova explosion then the maximum possible
velocity of the released star is $v_\mathrm{max}=\frac{3}{2}\;v_\mathrm{o}$.

If on the other hand, the companion of the massive star is a low-mass star 
then the ejection velocity of the binary is equal to the ejection velocity 
of the single star, thus we have $v_\mathrm{e}=v_\mathrm{o}$. The maximum possible
velocity of the released (less-massive companion) star 
after binary disintegration is of the order
$v_\mathrm{max}=2\;v_\mathrm{o}$. It might be expected  that in the case of 
binaries consisting of a high-mass and a low-mass star 
encountered by a high-mass
star the low-mass star will be ejected and an equal-mass binary will form. 
To what amount binaries with a large mass ratio will not suffer an 
exchange has to be 
quantified numerically in further studies.
 
\section{Conclusion}

Various theoretical and numerical studies exist on individual
ejection processes of massive stars, namely
the dynamical ejection scenario and the supernova ejection
scenario, but the combined effect of both scenarios for the 
distribution of massive stars in a galaxy 
has not been 
considered yet. 
In this paper we investigate for the first time 
the implications of the combination of the 
dynamical and the supernova ejection scenario for the O-star population 
of the Galactic field. We call this combined effect
the \emph{two-step-ejection} process of massive stars.
The main results are as follows:

i) The compound velocity, $v_\mathrm{c}$, which is the vectorial sum of the  
ejection
velocity, $v_\mathrm{e}$, of the binary from the star cluster and the 
release velocity, $v_\mathrm{r}$, with which
a star is released during a supernova, can be larger or smaller than
the previous ejection velocity. Stars can be both, accelerated and decelerated.

ii) The mean compound velocity is always greater than each of the 
initial velocities, $v_\mathrm{e}$ and $v_\mathrm{r}$ (eq.~\ref{eq_mean_v}).

iii) It is very unlikely that the parent star cluster of a massive field 
star produced by the two-step-ejection scenario can be identified. The
expected number fraction of such massive field stars which are formed apparently
in isolation can account quantitatively for the number of candidates 
for isolated massive star formation derived in \citet{dewit2005a}.

iv) Massive stars which are ejected via the two-step scenario can get higher
maximum space velocities (up to 1.5 times higher for equal-mass binary 
components or 2 times higher for significantly unequal-mass companion masses) 
than can be obtained by each process (dynamical or supernova ejection) 
individually.  

\bibliographystyle{mn2e}
\bibliography{OB-star,imf,star-formation,galaxy-evolution,cmf,star-cluster,n-body,onc,sn}

\begin{thebibliography}{}

\bibitem[\protect\citeauthoryear{{Blaauw}}{{Blaauw}}{1961}]{blaauw1961a}
{Blaauw} A.,  1961, \bain, 15, 265

\bibitem[\protect\citeauthoryear{{Chevalier}}{{Chevalier}}{1974}]{chevalier197%
4a}
{Chevalier} R.~A.,  1974, \apj, 188, 501

\bibitem[\protect\citeauthoryear{{Clarke} \& {Pringle}}{{Clarke} \&
  {Pringle}}{1992}]{clarke1992a}
{Clarke} C.~J.,  {Pringle} J.~E.,  1992, \mnras, 255, 423

\bibitem[\protect\citeauthoryear{{de Wit}, {Testi}, {Palla}, {Vanzi} \&
  {Zinnecker}}{{de Wit} et~al.}{2004}]{dewit2004a}
{de Wit} W.~J.,  {Testi} L.,  {Palla} F.,  {Vanzi} L.,    {Zinnecker} H.,
  2004, \aap, 425, 937

\bibitem[\protect\citeauthoryear{{de Wit}, {Testi}, {Palla} \& {Zinnecker}}{{de
  Wit} et~al.}{2005}]{dewit2005a}
{de Wit} W.~J.,  {Testi} L.,  {Palla} F.,    {Zinnecker} H.,  2005, \aap, 437,
  247

\bibitem[\protect\citeauthoryear{{Gies} \& {Bolton}}{{Gies} \&
  {Bolton}}{1986}]{gies1986a}
{Gies} D.~R.,  {Bolton} C.~T.,  1986, \apjs, 61, 419

\bibitem[\protect\citeauthoryear{{Gvaramadze} \& {Bomans}}{{Gvaramadze} \&
  {Bomans}}{2008}]{gvaramadze2008a}
{Gvaramadze} V.~V.,  {Bomans} D.~J.,  2008, \aap, 490, 1071

\bibitem[\protect\citeauthoryear{{Gvaramadze}, {Gualandris} \& {Portegies
  Zwart}}{{Gvaramadze} et~al.}{2008}]{gvaramadze2008b}
{Gvaramadze} V.~V.,  {Gualandris} A.,    {Portegies Zwart} S.,  2008, \mnras,
  385, 929

\bibitem[\protect\citeauthoryear{{Heggie}}{{Heggie}}{1980}]{heggie1980a}
{Heggie} D.~C.,  1980, in {D.~Hanes \& B.~Madore} ed., Globular Clusters
  {Dynamical theory of binaries in clusters}.
pp 281--+

\bibitem[\protect\citeauthoryear{{Hoffer}}{{Hoffer}}{1983}]{hoffer1983a}
{Hoffer} J.~B.,  1983, \aj, 88, 1420

\bibitem[\protect\citeauthoryear{{Hoogerwerf}, {de Bruijne} \& {de
  Zeeuw}}{{Hoogerwerf} et~al.}{2001}]{hoogerwerf2001a}
{Hoogerwerf} R.,  {de Bruijne} J.~H.~J.,    {de Zeeuw} P.~T.,  2001, \aap, 365,
  49

\bibitem[\protect\citeauthoryear{{Iben} \& {Tutukov}}{{Iben} \&
  {Tutukov}}{1997}]{iben1997a}
{Iben} I.~J.,  {Tutukov} A.~V.,  1997, \apj, 491, 303

\bibitem[\protect\citeauthoryear{{Kroupa}, {Aarseth} \& {Hurley}}{{Kroupa}
  et~al.}{2001}]{kroupa2001b}
{Kroupa} P.,  {Aarseth} S.,    {Hurley} J.,  2001, \mnras, 321, 699

\bibitem[\protect\citeauthoryear{{Lada} \& {Lada}}{{Lada} \&
  {Lada}}{2003}]{lada2003a}
{Lada} C.~J.,  {Lada} E.~A.,  2003, \araa, 41, 57

\bibitem[\protect\citeauthoryear{{Leonard}}{{Leonard}}{1991}]{leonard1991a}
{Leonard} P.~J.~T.,  1991, \aj, 101, 562

\bibitem[\protect\citeauthoryear{{Mikkola}}{{Mikkola}}{1983}]{mikkola1983a}
{Mikkola} S.,  1983, \mnras, 203, 1107

\bibitem[\protect\citeauthoryear{{Mikkola}}{{Mikkola}}{1984}]{mikkola1984a}
{Mikkola} S.,  1984, \mnras, 207, 115

\bibitem[\protect\citeauthoryear{{Parker} \& {Goodwin}}{{Parker} \&
  {Goodwin}}{2007}]{parker2007a}
{Parker} R.~J.,  {Goodwin} S.~P.,  2007, \mnras, 380, 1271

\bibitem[\protect\citeauthoryear{{Pflamm-Altenburg} \&
  {Kroupa}}{{Pflamm-Altenburg} \& {Kroupa}}{2006}]{pflamm-altenburg2006a}
{Pflamm-Altenburg} J.,  {Kroupa} P.,  2006, \mnras, 373, 295

\bibitem[\protect\citeauthoryear{{Portegies Zwart}}{{Portegies
  Zwart}}{2000}]{portegies_zwart2000a}
{Portegies Zwart} S.~F.,  2000, \apj, 544, 437

\bibitem[\protect\citeauthoryear{{Poveda}, {Ruiz} \& {Allen}}{{Poveda}
  et~al.}{1967}]{poveda1967a}
{Poveda} A.,  {Ruiz} J.,    {Allen} C.,  1967, Boletin de los Observatorios
  Tonantzintla y Tacubaya, 4, 86

\bibitem[\protect\citeauthoryear{{Stone}}{{Stone}}{1991}]{stone1991a}
{Stone} R.~C.,  1991, \aj, 102, 333

\bibitem[\protect\citeauthoryear{{Tauris} \& {Takens}}{{Tauris} \&
  {Takens}}{1998}]{tauris1998a}
{Tauris} T.~M.,  {Takens} R.~J.,  1998, \aap, 330, 1047

\bibitem[\protect\citeauthoryear{{Zwicky}}{{Zwicky}}{1957}]{zwicky1957a}
{Zwicky} F.,  1957, {Morphological astronomy}.
Berlin: Springer, 1957

\end{thebibliography}

\end{document}